\documentclass{article}

\usepackage{PRIMEarxiv}
\usepackage[utf8]{inputenc}
\usepackage[T1]{fontenc}
\usepackage{hyperref}
\usepackage{url}
\usepackage{booktabs}
\usepackage{amsfonts}
\usepackage{nicefrac}
\usepackage{microtype}
\usepackage{lipsum}
\usepackage{fancyhdr}
\usepackage{graphicx}
\usepackage{authblk}
\usepackage{amssymb}
\usepackage{xcolor}
\usepackage{epsfig}
\usepackage{subcaption}
\usepackage[section]{placeins}
\usepackage{mathrsfs}
\usepackage{graphicx}
\usepackage[notrig]{physics}
\usepackage{siunitx}
\usepackage{tikz} 
\usepackage[capitalise]{cleveref}

\title{A Simple Quantitative Model of Neuromodulation. \\ Part I: Ion Flow Through Neural Ion Channels}
\author[1]{Linda Werneck}
\author[2,3]{Mertcan Han}
\author[2]{Erdost Yildiz}
\author[1]{Marc-Andr\'e Keip}
\author[2,3,4]{\\ Metin Sitti}
\author[5,6,$\star$]{Michael Ortiz}

\affil[1]{{Institute of Applied Mechanics, University of Stuttgart, Stuttgart, Germany.}}
\affil[2]{{Physical Intelligence Department, Max Planck Institute for Intelligent Systems, Stuttgart, Germany.}}
\affil[3]{{Institute for Biomedical Engineering, ETH Zurich, Zurich, Switzerland.}}
\affil[4]{{School of Medicine and College of Engineering, Koc University, Istanbul, Turkey.}}
\affil[5]{{Division of Engineering and Applied Science, California Institute of Technology, Pasadena, California, USA.}}
\affil[6]{{Hausdorff Center for Mathematics, University of Bonn, Bonn, Germany.}}
\affil[$\star$]{Corresponding author: ortiz@aero.caltech.edu}
\DeclareRobustCommand{\Bd}{{\boldsymbol{\mathnormal d}}}
\DeclareRobustCommand{\Be}{{\boldsymbol{\mathnormal e}}}
\DeclareRobustCommand{\Bh}{{\boldsymbol{\mathnormal h}}}
\DeclareRobustCommand{\Bj}{{\boldsymbol{\mathnormal j}}}
\DeclareRobustCommand{\Bm}{{\boldsymbol{\mathnormal m}}}
\DeclareRobustCommand{\Bn}{{\boldsymbol{\mathnormal n}}}
\DeclareRobustCommand{\Bp}{{\boldsymbol{\mathnormal p}}}
\DeclareRobustCommand{\Bx}{{\boldsymbol{\mathnormal x}}}
\DeclareMathAlphabet{\Ibb}{U}{msb}{m}{n}
\DeclareRobustCommand{\IR}{{\Ibb R}}
\DeclareRobustCommand{\calB}{{\mathcal B}}
\DeclareRobustCommand{\calT}{{\mathcal T}}

\newcommand{\rmi}{\mathrm{i}}
\newcommand{\jumpl}{{[\kern-.15em[}}
\newcommand{\jumpr}{{]\kern-.15em]}}
\newcommand{\AND}{\quad\mbox{and}\quad}

\begin{document}

\maketitle

\begin{abstract}
We develop a simple model of ionic current through neuronal membranes as a function of membrane potential and extracellular ion concentration. The model combines a simplified Poisson-Nernst-Planck (PNP) model of ion transport through individual mechanosensitive ion channels with channel activation functions calibrated from {\sl ad hoc} in-house experimental data. The simplified PNP model is validated against bacterial Gramicidin A ion channel data. The calibrated model accounts for the transport of calcium, sodium, potassium, and chloride and exhibits remarkable agreement with the experimentally measured current-voltage curves for the differentiated human neural cells. All relevant data and code related to the ion flow models are available at \cite{darus}.
\end{abstract}

\section{Introduction}

Owing to its ability to provide non-invasive control of neural activity in deep-brain regions with millimeter spatial precision, ultrasonic neuromodulation (UNM) has elicited sustained interest (cf., e.~g., \cite{tyler2008a, tyler2010, tufail2010a, tyler2011a, tufail2011a, yoo2011a, younan2013a, mehic2014a, kamimura2016a, ye2016a, sato2018a, yoo2022}) and is widely regarded as one of the most significant new technologies for human neuroscience. Studies dating back to the 1950s \cite{wall1951a, barnard1955a, fry1958a, fry1958b} suggested that ultrasound can affect neural activity. In 2008, Tyler {\sl et al.} \cite{tyler2008a} reignited interest in this phenomenon by demonstrating neuromodulation in rodents using low-intensity ultrasound \cite{tyler2008a, tufail2010a} without a chemical or genetic pre-treatment of the subject. UNM complements human imaging techniques for studying brain connectivity and function in basic and clinical applications. Thus, established non-invasive modulation techniques such as transcranial magnetic and electrical stimulation (TMS and TES) are limited by their physics to mostly cortical regions and centimeter-scale resolution, thereby lacking access to subcortical areas underlying many neurological functions \cite{polania2018studying}. In contrast, the physics of ultrasound enables this modality to target deep tissue structures with millimeter precision, including the human brain \cite{bystritsky2011a, elias2016a}, which enables a broad range of applications, including drug delivery \cite{meng2021applications}, deep brain stimulation \cite{folloni2019a} and the treatment of epilepsy \cite{chen2019e}, and depression \cite{zhu2022transcranial}, among others.

Despite this surge in interest, the precise biophysical mechanisms underlying UNM have been the subject of extensive surmise and controversy. An unambiguous and definitive identification of such mechanisms has been finally effected by Yoo {\sl et al.} \cite{yoo2022}, who have shown that low-intensity focused ultrasound (LIFUS) in the 300--1000 kHz frequency range excites neurons through primarily mechanical means mediated by specific calcium-selective mechanosensitive (MS) ion channels. The activation of these channels results in a gradual build-up of calcium, which is amplified by calcium and voltage-gated channels, generating a burst firing response. Pharmacological and genetic inhibition of specific ion channels leads to reduced responses to ultrasound while over-expressing these channels results in stronger ultrasonic stimulation. In addition, Yoo {\sl et al.} \cite{yoo2022} find that cavitation \cite{bailey1996, plaksin2014a, plaksin2016a}, temperature changes \cite{tsui2005, shapiro2012a, constans2018a}, indirect auditory mechanisms \cite{sato2018a,guo2023}, and synaptic transmission are not required for this excitation to occur, thus ruling out other possible competing mechanisms. These findings strongly suggest the interaction between ultrasound and MS ion channels as the underlying mechanism responsible for UNM.

These remarkable contributions notwithstanding, a {\sl validated quantitative model} of the biophysical mechanisms by which LIFUS mechanically excites cortical neurons appears to be as yet unavailable. The overarching goal of the present work is to develop---and validate with in-house experiments on differentiated human neural cells---a {\sl simple} but predictive mechano-electro-physiological model of the effect of LIFUS on human cortical axons. The model, alongside imaging and characterization techniques such as functional magnetic resonance imaging (fMRI), magnetic resonance elastography (MRE) \cite{verhagen2019a, Legon12, Kamimura20, ophir1991a, nightingale2003a, Weickenmeier:2018, Lan:2020} and advanced computational modeling \cite{Budday:2020, salahshoor2020a}, can be used for medical device design and the optimization of personalized clinical procedures. We specifically aim to develop a multiscale hierarchy of electro-mechanical models that provide a fundamental understanding, as well as a quantitative and predictive capability, of how ultrasonic excitation interferes with brain activity and induces neuromodulation. 

At the full cranial scale, ultrasound wave propagation in the brain can be studied using finite element models representing a variety of conditions, from LIFUS to concussion (cf. e.~g.~\cite{sayed2008a, salahshoor2020a}). Detailed computational models have been successfully constructed from magnetic resonance (MR) images \cite{ganpule2017a, weickenmeier2017a, lu2019a}. The constitutive modeling of soft biological tissues has also received considerable attention \cite{sayed2008a, sayed2008b, mihai2015a, budday2017a, budday2017b, budday2019a}. These full-cranium computational models enable the precise determination of the viscoelastic wave patterns that arise in the human brain in response to LIFUS. In particular, for transducers operating at a fixed frequency, the models predict the steady-state harmonic deformations sustained by any target point within the brain or in any other organ of interest, such as the inner ear \cite{salahshoor2020a}.

Bianchi {\sl et al.} \cite{bianchi2018a, bianchi2019a} have shown that local strains such as those induced by LIFUS are transferred from tissue to individual cells in peripheral nerves. Specifically, they have quantified experimentally the changes in inwards and outwards ion currents and action potential (AP) firing in dorsal root ganglion‐derived neurons subject to uniaxial strains, using a custom‐built device allowing simultaneous cell deformation and patch clamp recording. In turn, the forces that regulate MS ion channel gating originate from the surrounding lipid membrane, suggesting a close relationship between membrane strain and MS ion channel function \cite{lundb1996a}. For example, membrane stiffening by stomatin-like protein‐3 has been shown to control mechanically gated ion channel activity in sensory neurons finely \cite{qi2015a}. In addition, voltage‐activated sodium channels have been shown to respond to strain, with a left‐shift in channel current-voltage (I–V) relations, leading to reduced inactivation potential, sodium leakage, and a reduced rate of AP firing \cite{rabut2020a}. Overall, these results suggest that ion channels are actuated directly by uniaxial strain in neuronal axons, thereby inducing cell electrophysiological activity.

The central question that remains is, therefore, how the electrophysiology of single neurons is influenced by axonal strain. We specifically aim to characterize this effect within the framework of the Hodgkin-Huxley (HH) model \cite{hodgkin1952a, hodgkin1952b, hodgkin1952c, hodgkin1952d}. In 1952, Alan Hodgkin and Andrew Huxley proposed their celebrated model to explain the ionic mechanisms underlying the initiation and propagation of action potentials in the squid giant axon, for which they received the 1963 Nobel Prize in Physiology or Medicine. In the HH model, voltage-gated ion channels are represented by effective electrical conductances depending on both voltage and time. The electro-chemical gradients driving the flow of ions through the channels are represented by sources whose voltages are determined by the ratio of the intra- and extracellular concentrations of the ionic species under consideration. 

Within this framework, we posit that a quantitative model of mechanical neuromodulation can be fashioned in three steps: i) a model of conductance due to ion flow through open axonal channels; ii) a model of mechanosensitive channel actuation by a prescribed axial strain \cite{boland2008, tyler2008a, sachs2010}; and iii) a model of parametric resonance resulting from axonal harmonic excitation such as induced by LIFUS. The validation of these three elements of the model require specialized experimental protocols and extensive laboratory testing. Therefore, we divide the presentation of the model into three parts. In the present part I, we focus on the effective ionic current of ion channels and their dependence on ion concentration and channel geometry, which lays the foundation for the remaining parts II and III of the model, to be presented in subsequent publications. 

We specifically model ion transport by recourse to the coupled Poisson-Nernst-Planck (PNP) equations \cite{fogolari1997a, zheng2011a}. The basic framework of the PNP model and some of its extensions, including size effects \cite{eisenberg2011a}, ion-water interactions \cite{chen2016b}, coupling to density functional theory \cite{meng2014a}, and others, is presently well-established. However, the direct first-principles or molecular dynamics simulation of mechanical gating in ion channels remains largely out of reach due to the structural complexity of the channels, disparate length scales, and the staggering gap between the molecular and diffusive time scales. Present approaches are, for the most part, decoupled from the mechanical response and treat the surface and charge distribution of the channels as given. 

A coarse-grained strategy for ion channel analysis is to construct a continuum functional of a charge transport system to encompass the polar and nonpolar free energies of solvation and chemical potential-related energies expressed in terms of averaged ion concentrations \cite{chen2016a}. Using the calculus of variations, a coupled PNP system of equations and other transport equations then follows whose solutions give explicit profiles of electrostatic potential and densities and fluxes of charged species \cite{fogolari1997a, zheng2011a}. 

Our modeling strategy consists of solving the PNP equations at the continuum level for individual ion channels assuming a simplified cylindrical channel geometry, and then estimating the membrane conductances arising in the HH model by means of a mean-field approximation that uses the density of channels per unit area of the axonal membrane. These simplifications set forth a simple quantitative model of ion flow that characterizes the effective conductance of the axonal membrane and parametric dependencies thereof, including channel geometry. The model accounts for the effect of calcium, sodium, potassium, and chloride ions. 

Evidently, the critical question to be addressed is whether such a simple model suffices to characterize effective ion conductances accurately. To elucidate this question, we validate the model against archival experimental data from a single bacterial ion-channel model, gramicidin A, and calibrate it with in-house experimental data from a human neural cell culture acquired through electrophysiological recordings conducted specifically for the present study. We find that the agreement between the predictions of the continuum model and the experimental data is excellent, which is remarkable considering the simplicity of the model. This validation suggests that the effective ion conductance of axonal membranes depends mainly on coarse-grained channel parameters such as cross-sectional area and channel length and, to a good approximation, is independent of the fine structure of the channels.

In Section \ref{sec:2ionflow} we first introduce the PNP equations in form of a thermodynamic framework. Based on the PNP, we present a model for ion flow through single ion channels and we validate our results using data from GA channels in Section \ref{sec:3singlechannel}. We extend the single channel model to a full-axon model using in-house experimental data and discuss our results in Section \ref{sec:4fullaxon}. Finally, we conclude our work in Section \ref{sec:5conclusion}.

\section{A simple quantitative model of ion flow through axonal channels}
\label{sec:2ionflow}

Given the crucial role played by mechanosensitive (MS) ion channels in the physiology of mechanotransduction, considerable effort has been devoted to understanding their gating mechanisms (cf.~\cite{martinac2004a} for a review). MS ion channels respond to mechanical forces along the plane of the cell membrane (membrane tension). The protein structure of many MS ion channels is known and available in protein repositories (GenBank, Protein DataBank, and SwissProt), which provides a basis for molecular dynamics simulations. MS ion channels of small conductance (MscS) from several prokaryotes have been extensively characterized \cite{perozo2002a} and serve as model systems for understanding the physio-chemical principles of mechanotransduction. Comparatively, much less is known about the structure of eukaryotic members of the MscS superfamily, many of which acquire extra transmembrane helices as well as additional extra-membrane domains \cite{deng2020a}, but the number of newly characterized channels is growing at a rapid pace \cite{rnad2010a, ranade2015a}. Even in cases where the protein structure is known in detail, the first-principles characterization of the gating mechanism of MS ion channels, with or without strain, remains a formidable challenge. An additional challenge in understanding the function of MS ion channels concerns the analysis of ion transport. Ion channel exists in a complex environment, including cell membrane, water molecules, mobile ions, and other molecular components. These components interact through mutual long-range (e.~g., electrostatics) and short-range (e.~g., Lennard-Jones) interactions.

Unlike these first-principles studies, the focus of the present work is to estimate the effective ionic conductance of axonal membranes as a function of coarse features of the channels, such as size and density, with a view to devising a strain- and vibration-dependent HH model of neuromodulation. To this end, we resort to a continuum PNP model \cite{fogolari1997a, zheng2011a, chen2016a} applied to simplified cylindrical channel geometry and subsequently estimate the membrane conductances by means of a mean-field approximation based on channel densities. The PNP model is a mean-field model that, using a continuum approximation, treats the ion flow as the averaged ion concentration driven by the electrostatic potential force and ion concentration gradient. These aspects of the models are reviewed next for completeness and ease of reference.

\subsection{Thermodynamic framework}

We consider a body in a configuration $\calB \subset \IR^3$ with boundary $\partial \calB$ at a given time $t$ in a time interval $\cal T \subset \IR_+$. The body is contained in a domain $\Omega \subset \IR^3$ in free space, which includes the space occupied by the body $\calB$. The space is spatially parameterized in the coordinates $\Bx \in \Omega$. To describe electro-chemical phenomena, we introduce as independent field variables the electric potential
\begin{equation}
    \phi: 
    \begin{cases} 
        \begin{array}{l}
            \calB \times \calT \rightarrow \IR \\
            (\Bx,t) \mapsto \phi (\Bx,t)
        \end{array}
    \end{cases}
    \label{eq:primary_electric}
\end{equation}
and the concentrations of the individual species, labeled by '$\rmi$', as well as the corresponding chemical potentials,
\begin{equation}
    c_\rmi:
    \begin{cases} 
        \begin{array}{l}
            \calB \times \calT \rightarrow \IR \in [0,1] \\
            (\Bx,t) \mapsto c_\rmi (\Bx,t)
        \end{array}
    \end{cases}
    \AND
    \mu_\rmi:
    \begin{cases}
        \begin{array}{l}
            \calB \times \calT \rightarrow \IR \\
            (\Bx,t) \mapsto \mu_\rmi (\Bx,t) .
        \end{array}
    \end{cases}
    \label{eq:primary_chemical}
\end{equation}
The driving electro-chemical fields are the electric field $\Be$ and the chemical fields $\Bm_\rmi$, defined as
\begin{equation}
    \Be := - \nabla \phi
    \AND
    \Bm_\rmi := - \nabla \mu_\rmi ,
    \label{eq:gradient_fields}
\end{equation}
where `$\nabla$' represents the gradient operator with respect to $\Bx$. An application of Cauchy's theorem further defines the electric displacement field $\Bd$ and the molar ion-flux densities $\Bh_\rmi$, with associated jump conditions
\begin{equation}
    - q_{\rm f} = \jumpl \Bd \jumpr \cdot \Bn
    \AND
    0 = \jumpl \Bh_\rmi \jumpr \cdot \Bn ,
    \label{eq:secondary}
\end{equation}
where $q_{\rm f}$ is the surface density of free electric charges. In the above expressions, $\jumpl \bullet \jumpr$ denotes the jump of a quantity $\bullet$ across a surface $\cal S$ separating two regions $(1)$ and $(2)$ such that $\jumpl \bullet \jumpr := \bullet_{(1)} - \bullet_{(2)}$ and $\Bn$ denotes a unit normal on $\cal S$ pointing from the region (1) to the region (2). \cref{eq:secondary}$_2$ describes continuous molar flux across surfaces, where the molar surface-flux densities---characterizing the amount of ion species passing through a given surface per unit time---are given as $h_\rmi := \Bh_\rmi \cdot \Bn$. The electric displacement can be expressed in terms of the polarization $\Bp$ as
\begin{equation}
    \Bd = \epsilon_0 \Be + \Bp ,
    \label{eq:d_p}
\end{equation}
where $\epsilon_0~\approx~8.854~\times~10^{-12}~\frac{\textnormal{F}}{\textnormal{m}}$ is the electric permittivity of free space and the polarization $\Bp$ only exists in space filled with polarizable matter. 

The motion of the ions induces ion-current densities
\begin{equation}
    \Bj_\rmi = F Z_\rmi \Bh_\rmi ,
\end{equation}
where $Z_\rmi$ is the ionic valence of the individual species and $F~\approx~9.6485~\times~10^4$ $\frac{\textnormal{C}}{\textnormal{mol}}$ is the Faraday constant. The total ion-current density is then
\begin{equation}
    \Bj = \sum_\rmi \Bj_\rmi ,
\end{equation}
with associated jump condition
\begin{equation}
    0 = \jumpl \Bj \jumpr \cdot \Bn ,
\end{equation}
where $j := \Bj \cdot \Bn$ is the surface density of ionic current. We assume throughout weak electric currents, from which no magnetic fields of significant magnitude are created.

\subsection{Balance equations and dissipation inequality}

Gauss's law of electrostatics requires
\begin{equation}
    \mbox{div} \, \Bd = \rho ,
    \label{eq:gauss_law}
\end{equation}
where $\rho$ is the volumetric density of ionic charges, related to the molar concentrations $c_\rmi$ via
\begin{equation}
    \rho = F \sum_\rmi Z_\rmi c_\rmi .
    \label{eq:rho}
\end{equation}
In addition, assuming that a change in concentration within a control volume can only happen as a result of an inward or outward flux of matter through the surface of the control volume (and cannot be generated locally within the control volume), we obtain the mass balance laws for the ion concentrations
\begin{equation}
    \dot c_\rmi = - \mbox{div} \, \Bh_\rmi.
    \label{eq:mass_balances}
\end{equation}

We posit an additive decomposition of the electro-chemical energy-density function
\begin{equation}\label{eT9YOk}
    \widehat \Psi ( \Be, c ) 
    := 
    \widehat \Psi_0 ( \Be ) 
    + 
    \widehat \Psi_{\rm mat} ( \Be, c ) ,
\end{equation}
where we write $c := (c_1,\dots,c_n)$, with $n$ the number of ionic species, 
\begin{equation}
    \widehat \Psi_0 ( \Be)
    =
    - 
    \frac{\epsilon_0}{2} \| \Be \|^2 ,
\end{equation}
$\Bx \in \Omega$, is the electric energy density of free space and $\widehat \Psi_{\rm mat} ( \Be, c_\rmi )$, $\Bx \in \calB$, is the electro-chemical energy density in the presence of matter. The corresponding thermodynamically consistent relationships for the polarization and the chemical potential then follow as
\begin{equation}
    \Bp 
    := 
    - 
    \partial_\Be \widehat \Psi_{\rm mat} ( \Be, c ) ,
    \AND
    \mu_\rmi 
    := 
    \partial_{c_\rmi} \widehat \Psi_{\rm mat} ( \Be, c ) .
    \label{eq:mu_and_d}
\end{equation}
Using these relations, the dissipation inequality reduces to
\begin{equation}
    \sum_\rmi ( \Bm_\rmi + Z_\rmi F \Be ) \cdot \Bh_\rmi \ge 0 .
    \label{eq:reduced_dissipation_inequality}
\end{equation}
By further defining the electro-chemical potentials of the individual ions, as well as the corresponding negative gradients, as
\begin{equation}
    \overline{\mu}_\rmi := \mu_\rmi + Z_\rmi F \phi
    \AND
    \overline{\Bm}_\rmi := - \nabla \overline{\mu}_\rmi
\end{equation}
and using the definition of $\Bm_\rmi$ and $\Be$ as gradient fields according to \eqref{eq:gradient_fields}, we recast the dissipation inequality \eqref{eq:reduced_dissipation_inequality} in the compact form 
\begin{equation}
    \sum_\rmi \overline{\Bm}_\rmi \cdot \Bh_\rmi \ge 0 .
\end{equation}
This constraint can be automatically fulfilled by assuming kinetic relations for the molar flux densities of the individual ions of the potential form
\begin{equation}
    \Bh_\rmi 
    := 
    \partial_{\overline{\Bm}_\rmi} \widehat \Phi_\rmi ( \overline{\Bm}_\rmi ) ,
    \label{eq:h}
\end{equation}
and requiring the electro-chemical dissipation-potential-density function $\widehat \Phi_\rmi ( \overline{\Bm}_\rmi )$ to be a {\sl gauge}, i.~e., a non-negative, positively homogeneous and convex function that evaluates to zero at the origin.

\subsection{Material model}

We model electro-chemical (electrodiffusive) behavior by recourse to the classical PNP equations. In this setting, the electro-chemical energy-density function of the material is assumed to be of the form
\begin{equation}
   \widehat \Psi_{\rm mat} ( \Be, c )
    := 
    - 
    \dfrac{\epsilon_0 \chi}{2} \| \Be \|^2
    + 
    \sum_{\rmi=1}^n
    R T c_\rmi 
    \left[ 
        \mbox{log} \frac{c_\rmi}{c_0} 
        - 
        \left(1 - \frac{c_0}{c_\rmi} \right) 
    \right] , 
\end{equation}
where $\chi$ is the electric susceptibility, $R \approx 8.314 $ $\frac{\textnormal{J}}{\textnormal{mol K}}$ is the gas constant, $T$ is the absolute temperature and $c_0$ is the reference molar concentration. In addition, the dissipation-potential-density functions are assumed to be of the form
\begin{equation}
   \widehat \Phi_\rmi ( \overline{\Bm}_\rmi)
    := 
    \frac{1}{2} \frac{D_\rmi}{R T} c_\rmi \| \overline{\Bm}_\rmi \|^2 ,
\end{equation}
where $D_\rmi$ are the diffusion coefficients of the ionic species.

For this particular model, an application of \eqref{eq:mu_and_d} gives the electric polarization and the chemicals potentials as
\begin{equation}
    \Bp = \epsilon_0 \chi \Be
    \AND
    \mu_\rmi = RT \, \mbox{log} \frac{c_\rmi}{c_0} ,
    \label{eq:d_and_mu}
\end{equation}
whence the electric displacement $\Bd$ in the entire domain can be computed using \eqref{eq:d_p}. In addition, an application of \eqref{eq:h} gives the Nernst-Planck equation
\begin{equation}
    \Bh_\rmi 
    = 
    - 
    D_\rmi \left( \nabla c_\rmi 
    + 
    \frac{Z_\rmi F}{R T} \, c_\rmi \, \nabla \phi \right) .
\end{equation}
Inserting these relations into the conservation equations \eqref{eq:mass_balances}$_1$ gives the diffusion equations
\begin{equation}
    \dot c_\rmi 
    = 
    \mbox{div} 
    \left[ 
        D_\rmi 
        \left( 
            \nabla c_\rmi 
            + 
            \frac{Z_\rmi F}{R T} \, c_\rmi \, \nabla \phi 
        \right) 
    \right] .
    \label{eq:ge1}
\end{equation}
In addition, Gauss's law \eqref{eq:gauss_law} together with the electric displacement \eqref{eq:d_p}, the electric field \eqref{eq:gradient_fields}, and the relation \eqref{eq:rho} gives
\begin{equation}
    - 
    \mbox{div}
    ( 
        \epsilon \nabla \phi 
    ) 
    = 
    F \sum_\rmi Z_\rmi \, c_\rmi ,
    \label{eq:ge2}
\end{equation}
where $\epsilon = \epsilon_0 ( 1 + \chi)$ is the electric permittivity.

Eqs.~(\ref{eq:ge1}) and (\ref{eq:ge2}) set forth a system of partial differential equations which, together with suitable initial and boundary conditions, govern the evolution of the ionic concentrations and the electrostatic field. 

\section{Mass transport through single ion channel}
\label{sec:3singlechannel}

As a first step towards the formulation of a full-axon ion transport model, we begin by considering an individual ion channel in isolation and seek to characterize the net flux of ions through the channel by means of the PNP model described in the foregoing. 

\subsection{Cylindrically symmetric PNP problem}

\begin{figure}[hbt!]
\centering
\includegraphics{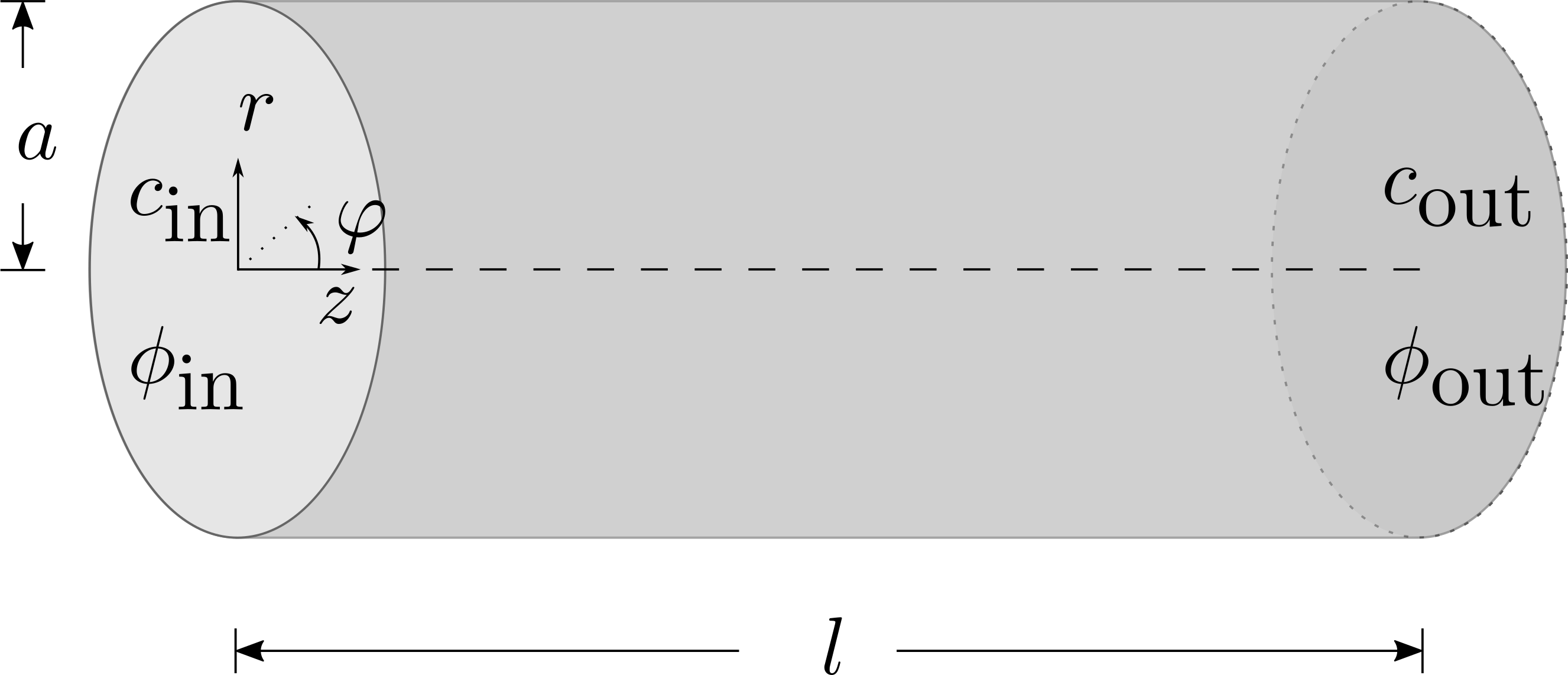}
\caption{Simplified geometry and boundary conditions assumed in calculations of ion channel mass transport.}
\label{fig:channelgeometry}
\end{figure}

In order to obtain a simple and easy-to-evaluate model, we assume a cylindrical ion-channel geometry with boundary conditions shown in \cref{fig:channelgeometry} and refer the solution to a system of cylindrical coordinates $(r,\theta,z)$. In this representation, the domain of analysis is $0 \leq r \leq a$, $0 \leq \theta < 2\pi$, $0 \leq z \leq l$, where $a$ and $l$ are the radius and length of the channel, respectively. We further assume that the lateral surface of the channel is charge-free and impermeable to the ions and that the electric permittivity of the channel is homogeneous. Finally, we assume that the ion concentration is at a steady state and uniform across cross-sections of the channel. 

By virtue of these simplifying assumptions, the concentration and electrostatic potential fields, $c(z)$ and $\phi(z)$, respectively, depend on the axial coordinate $z$ only, and the governing equations \eqref{eq:ge1} and \eqref{eq:ge2} reduce to the elementary form
\begin{equation}
    c^{\prime \prime}
    +
    \frac{Z F}{R T}
    \left( c \phi^{\prime} \right)^\prime
    =
    0 ,
    \qquad
    \epsilon
    \phi^{\prime \prime}
    +
    Z F c
    =
    0 ,
\end{equation}
respectively, where $(\cdot)^\prime$ denotes the spatial derivative with respect to the coordinate $z$. We further assume the boundary conditions
\begin{equation}
    c(0) = {c}_{\rm in} ,
    \quad
    c(l) = {c}_{\rm out} ,
    \quad
    \phi(0) = {\phi}_{\rm in},
    \quad
    \phi(l) = {\phi}_{\rm out} ,
\end{equation}
where ${c}_{\rm in}$ and ${\phi}_{\rm in}$ are the concentration and electrostatic potential at the inlet of the channel, respectively, with ${c}_{\rm out}$ and ${\phi}_{\rm out}$ {\sl idem} at the outlet. The instantaneous ion flux through the channel can be computed from the Gauss theorem as
\begin{equation}
    {I}_{\rm c} = Z F \int_{A_{\rm in/out}} D \left( c^\prime + \frac{Z F}{R T} c \phi^\prime\right)_{z=0} {\rm d} A ,
    \label{eq:channel_current}
\end{equation}
where $A_{\rm in/out}$ are the inlet/outlet sections of the channel, respectively.

\subsection{Validation: Gramicidin A channels}
\label{sec:res_channel}

\begin{figure}[hbt!]
\centering
\includegraphics[width=50mm,scale=0.33]{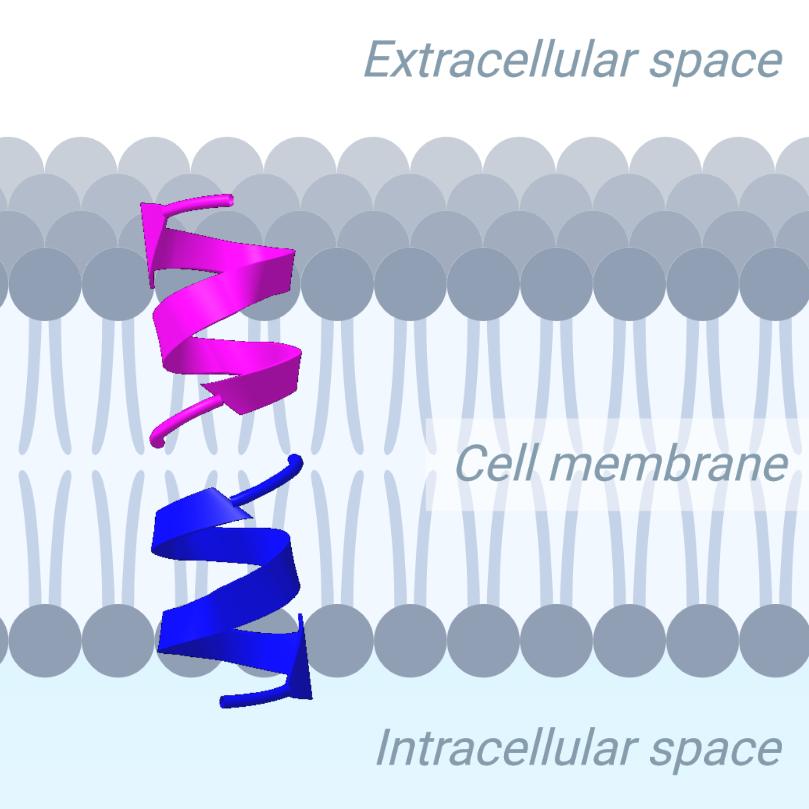}
\caption{Alignment of GA channel in the cellular membrane. The channel consists of 15 L- and D-amino acids and presents as helix dimers in the lipid bilayers.}
\label{fig:gramicidin}
\end{figure}

By way of validation of the single-channel model just described, we compare the predictions of the model against archival experimental data \cite{busath1998gramicidin, smart1993pore} for Gramicidin A (GA) channels in NaCl and KCl solutions at different molar concentrations, \cref{fig:gramicidin}. Extra and intracellular concentrations take equal values between 0.2 and 2~mM (molarity, 1 M is 1 mol/l), and the membrane potential varies between 25 and 200~mV. The properties of the GA channel used in the calculation are collected in \cref{tab:GA_properties}. We note that agreement with the observational data requires effective diffusion coefficients in GA channels to be lower than the bulk diffusion coefficients by factors of $2$ to $10$ \cite{mamonov2006gramicidin}.

\begin{table}[h]
\caption{Properties of GA channel \cite{dani1981gramicidin, mamonov2006gramicidin, smart1993pore} used in PNP calculations.}
\label{tab:GA_properties}
\centering
\begin{tabular}{c l}
\hline
Parameter & Value \\\hline
$l_\text{GA}$    & $2.60 \times 10^{-9} \, \textnormal{m}$   \\ 
$a_\text{GA}$    & $2.00 \times 10^{-10} \, \textnormal{m}$  \\
$D_\text{GA,Na}$ & $1.33 \times 10^{-10} \, \frac{\textnormal{s}}{\textnormal{m}^2}$  \\
$D_\text{GA,K}$  & $3.92 \times 10^{-10} \, \frac{\textnormal{s}}{\textnormal{m}^2}$   \\
$D_\text{GA,Cl}$ & $2.03 \times 10^{-10} \, \frac{\textnormal{s}}{\textnormal{m}^2}$  \\
\hline
\end{tabular}
\end{table}

\begin{figure}[htb!]
\centering
\includegraphics[width=\textwidth]{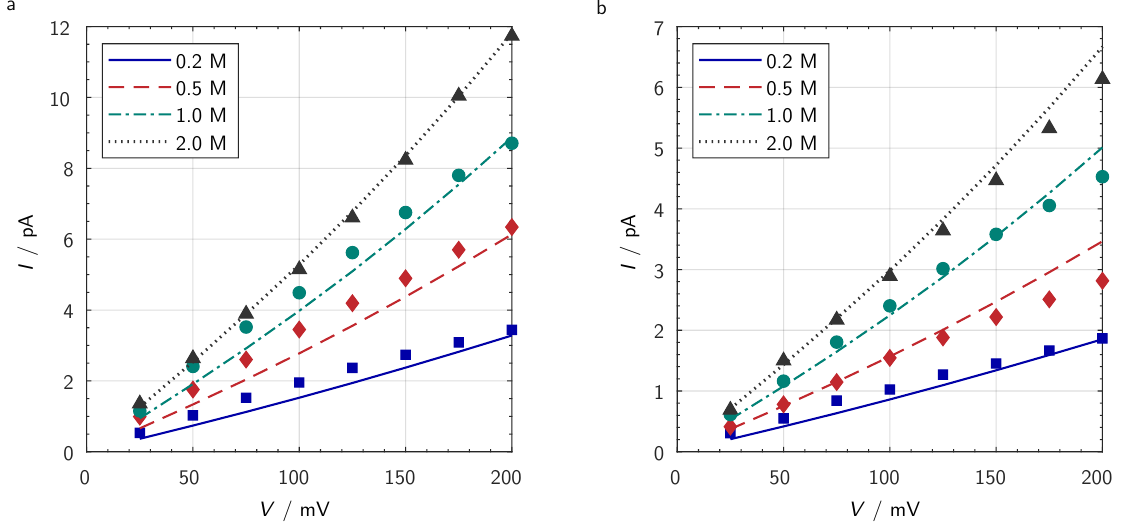}
\caption{Comparison between experimental data \cite{busath1998gramicidin} and simulation results of voltage clamp of (a) KCl, and (b) NaCl solution for different molar concentrations.}
\label{fig:GA}
\end{figure}

\cref{fig:GA} shows experimental data points and computed ionic currents through a single GA channel for different NaCl and KCl concentrations. The agreement between the model predictions and the experimental data is remarkable, especially in view of the simplicity of the model, which suggests that the effective ion transport of individual channels depends mainly on coarse channel parameters such as cross-sectional area and channel length and is independent of the fine structure of the channels to a good first approximation.

\section{Full-axon ion transport model}
\label{sec:4fullaxon}

Excitation and electrical signaling of neurons involve the transport of ions through channels that populate the neuronal membrane in large numbers \cite{Hille:2001}. The sodium (Na${}^+$), potassium (K${}^+$), calcium (Ca${}^{2+}$) and chloride (Cl${}^-$) ions account for the majority of the action. Each channel responds to a voltage, chemical or mechanical stimuli, and the response of the channel is called {\sl gating}, which keeps the channel open for a few milliseconds. The open channels exhibit selective permeability, allowing a specific species of ion to flow at high rates of the order of $10^6$ ions per second. Neurons exhibit a voltage $V$, i.~e.\ an electrical potential difference across their membrane, which is negative at the cytoplasmic side compared to the extracellular space. This membrane voltage, normally ranging from $-90$ to $-50$ mV, is mainly due to differences in ionic concentration at the extra- and intracellular sides of the membrane. In order to estimate the net ionic current through an entire axonal membrane, we resort to a simple mean-field model that aggregates the ion fluxes of the individual channels such as computed, e.~g., by the PNP model set forth in Section~\ref{sec:2ionflow}. 

\subsection{A mean-field model for voltage-gated channels} 

We consider voltage-gated Na${}^+$, K${}^+$, Ca${}^{2+}$ and Cl${}^-$ channels responding to a prescribed voltage $V$. We assume that the overall ion transport $I_\text{tot}$, through the membrane is the sum of the ion currents through all the individual channels that are open at a given voltage $V$. We posit that the number of gated channels is an increasing function of $V$, to be determined. These assumptions suggest a relation of the form
\begin{equation}
    I_\text{tot} 
    = 
    b_\text{Na}(V) {I}_{\rm c}^\text{Na} 
    + 
    b_\text{K}(V) {I}_{\rm c}^\text{K} 
    + 
    b_\text{Ca}(V) {I}_{\rm c}^\text{Ca} 
    + 
    b_\text{Cl}(V) {I}_{\rm c}^\text{Cl} ,
    \label{eq:linearCombination}
\end{equation}
where ${I}_{\rm c}^\text{Na}$, ${I}_{\rm c}^\text{K}$, ${I}_{\rm c}^\text{Ca}$ and ${I}_{\rm c}^\text{Cl}$ are ionic currents of single Na${}^+$, K${}^+$, Ca${}^{2+}$ and Cl${}^-$ channels, respectively. In addition, $b_\text{Na}(V)$, $b_\text{K}(V)$, $b_\text{Ca}(V)$ and $b_\text{Cl}(V)$ are the number of Na${}^+$, K${}^+$, Ca${}^{2+}$ and Cl${}^-$ channels in the axon, respectively, that are gated at voltage $V$. To close the model, we calibrate these activation functions empirically from our in-house experimental data, as described next.

\subsection{Experimental setup: Patch-clamp recordings from differentiated human neural cells}

\begin{figure}[ht]
     \centering
     \includegraphics[width=\textwidth]{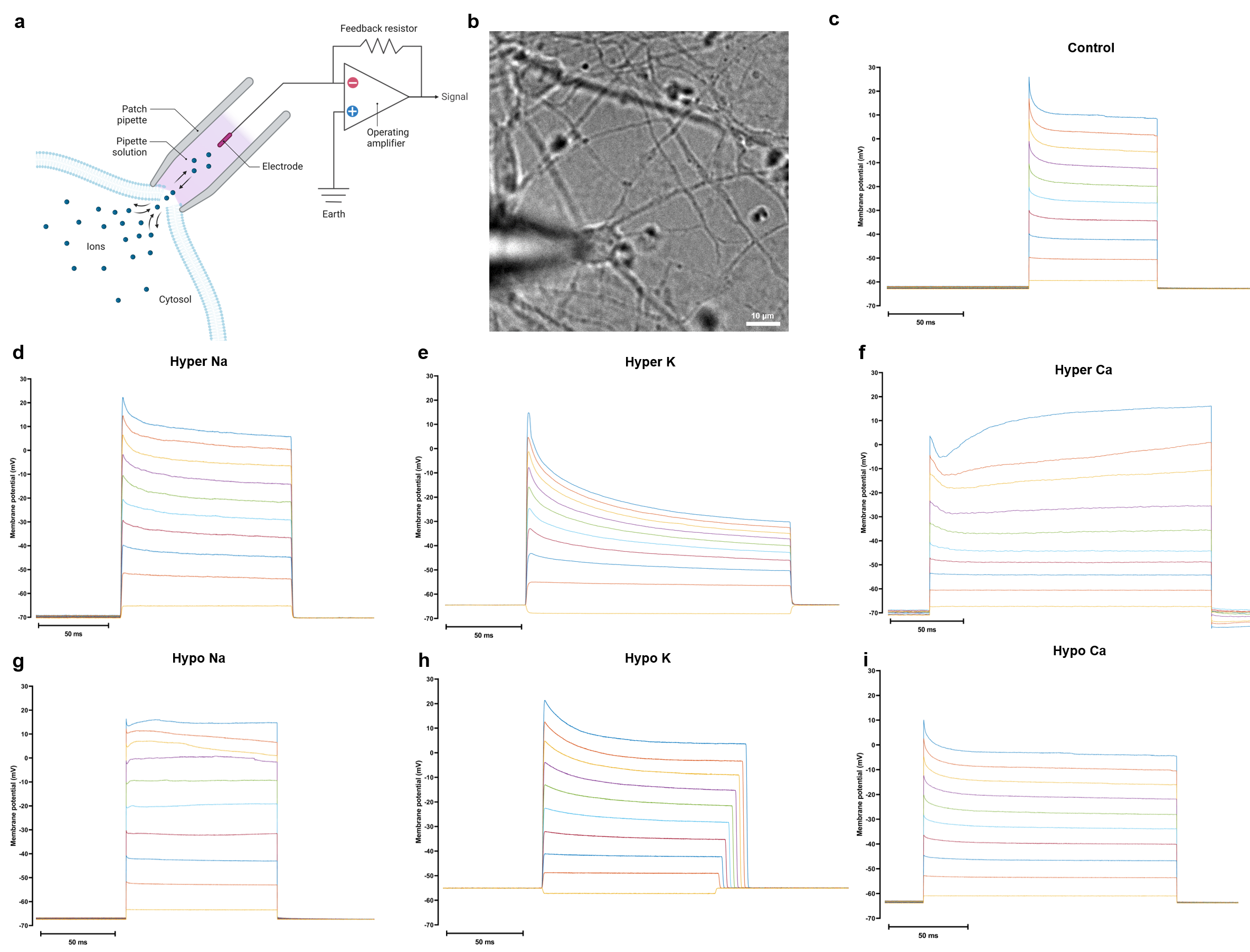}
     \caption{a) Experimental setup; b) Phase-contrast microscopy image during patch-clamp measurements; c--i) Representative traces of membrane voltage in response to step depolarization between $-70$ and 30~mV under various ion concentration conditions; c) Physiological ion concentrations; d) Hyperphysiological sodium concentration; e) hyperphysiological potassium concentration; f) hyperphysiological calcium concentration; g) hypophysiological sodium concentration; h) hypophysiological potassium concentration; i) hypophysiological calcium concentration.}
     \label{fig:expfigIV}
\end{figure}

For purposes of model calibration, we have tested a human-derived neural progenitor cell line (ReNcell CX, Sigma Aldrich, MO, USA). The neurons are maintained by means of ReNcell NSC maintenance media on laminin-coated cell culture dishes at \SI{37}{\degreeCelsius} in a humidified incubator with 5$\%$ CO$_2$ prior to use for experiments. After the differentiation of the neural cells, all electrophysiological recordings are made with a whole-cell patch clamp setup (Axopatch 200B, Molecular Devices, CA, USA). The pulled patch pipettes are utilized at $4$--$6$ M$\Omega$ resistance to carry out whole-cell patch-clamp experiments. The physiological extracellular media are prepared by mixing 20 mM HEPES, 10 mM glucose, 140 mM NaCl, 2.5 mM KCl, 1.8 mM CaCl$_2$, and 1.0 mM $\text{MgCl}_2$ in distilled water. The pH was calibrated to 7.4 using 1 M NaOH. The ion concentrations for other extracellular media conditions are modified according to \cref{tab:ExpIonCon}. The internal cellular medium is purchased by a commercial producer (Internal KF 110, Nanion, Munich, Germany). For whole-cell patch-clamp measurements, the dynamic current-voltage measurements are recorded from the same point in the axon hillock of each neuron for standardization of the experimental data collection. The patch pipettes are filled with the intracellular solution to achieve a whole-cell patch, and pipette tips are applied to cells while holding positive pressure. When whole-cell patch formation is achieved after gigaseal is overpassed, the patch-clamp pipette is held in this position for 5 minutes to stabilize the membrane potential in the physiological range. Dynamic current-voltage measurements are made from $-100$ to 100~mV with 10~mV steps in the voltage clamp settings. Representative experimental results for membrane potential in response to step depolarization between the physiological range of membrane potential are shown in \cref{fig:expfigIV}.

\begin{table}[hbt!]
\caption{Extracellular ion concentrations (mM) in modified and physiological neuronal media.}
\label{tab:ExpIonCon}
\centering

\begin{tabular}{l l l l}
\hline
Ion & Hypophys. & Phys. & Hyperphys. \\\hline
$c_\text{out}^\text{Na}$ & $120$ & $140$ & $160$ \\
$c_\text{out}^\text{K}$ & $1.5$ & $2.5$ & $6.0$\\
$c_\text{out}^\text{Ca}$ & $1.0$ & $1.8$ & $4.0$ \\
\hline
\end{tabular}

\end{table}

\subsection{Calibration of an ionic transport model for neuronal membranes} 
\label{sec:res_membrane}

\begin{figure}[ht]
     \centering
     \includegraphics[width = 0.5\textwidth]{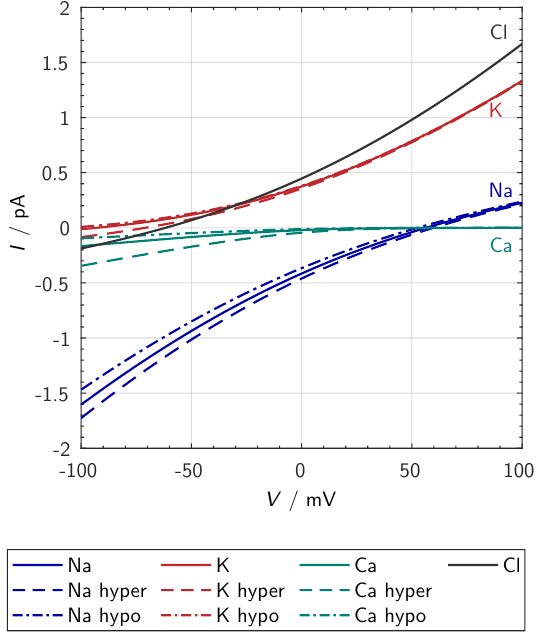}
     \caption{Single-channel currents for individual ions computed by PNP equations for physiological as well as hyper- and hypophysiological concentrations.}
     \label{fig:PNPcurrents}
\end{figure}

The boundary conditions of the PNP model are according to the experimental setup. In calculations, membrane potentials are varied between $-100$ and 100~mV. The evaluated extracellular concentrations are listed in \cref{tab:ExpIonCon}; intracellular concentrations are given by the physiological state of neurons as $c_\text{in}^\text{Na} = 15~\text{mM}$, $c_\text{in}^\text{K} = 100~\text{mM}$, $c_\text{in}^\text{Ca} = 2 \times 10^{-4}~\text{mM}$, $c_\text{in}^\text{Cl} = 13~\text{mM}$. Representative results from PNP calculations are shown in \cref{fig:PNPcurrents}.

For the physiological concentrations, the equilibrium potentials $E_\text{ion}$ for all ion species agree closely with the literature values $E_\text{Na} = 56.4~\text{ mV}$, $E_\text{K} = - 93.1~\text{mV}$, $E_\text{Ca} = 115.0~\text{mV}$, $E_\text{Cl} = -61.7~\text{mV}$ \cite{bear2007neuroscience}. In the hyperphysiological environment, the concentration gradients for sodium and calcium are increased, while the concentration gradient of potassium is decreased. These concentrations result in larger sodium and calcium currents and a smaller potassium current. Contrarily, in the hypophysiological scenario smaller concentration gradients of sodium and calcium result in smaller currents, whereas the potassium current is increased by a higher potassium concentration gradient. This effect is dominant for smaller membrane potentials and relatively minor for potassium compared with the other cations.

\begin{figure}[ht]
     \centering
     \includegraphics[width=\textwidth]{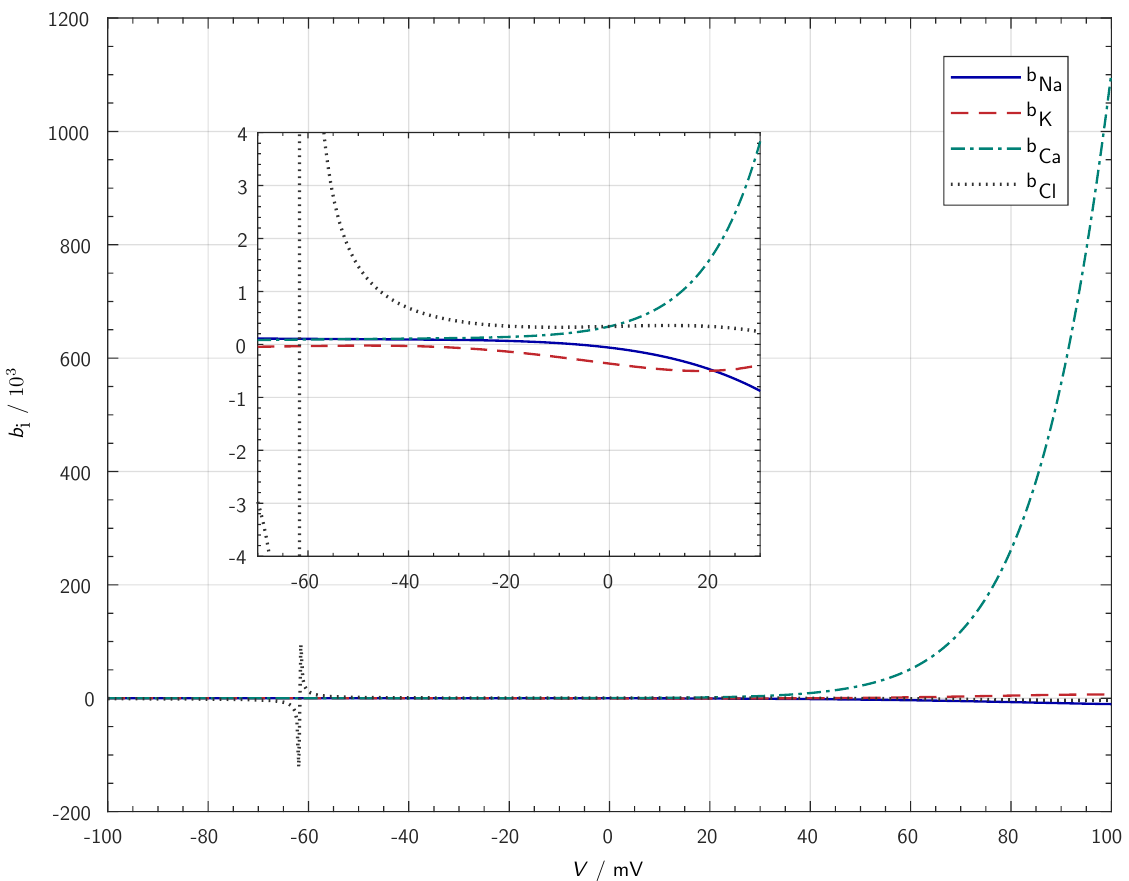}
     \caption{Activation parameters of sodium, potassium, calcium, and chloride ions in experimental and in physiological range.}
     \label{fig:parameters}
\end{figure}

With the single-channel currents computed by the PNP model, multiple linear regression of \cref{eq:linearCombination} to the data is carried out in order to identify the activation functions $b_\text{Na}(V)$, $b_\text{K}(V)$, $b_\text{Ca}(V)$ and $b_\text{Cl}(V)$ over the full range of the membrane potentials. The activation functions thus identified are shown in \cref{fig:parameters}. The activation functions encode the information contained in the data, ranging from the number of active channels at different membrane potentials, the relative activities of the ionic species and their interaction, and other information. 

Two features immediately stand out in \cref{fig:parameters}: i) the chloride activation function exhibits a discontinuous spike at the chloride resting potential, and ii) the activation function of calcium takes high values for membrane potentials $ > 40$ mV outside of the physiological range, as the membrane potential approaches the calcium equilibrium potential. Both activation functions stabilize the ionic current flow. This effect is expected for calcium channels in view of the strong influence that calcium ions have on the overall cell response, Fig.~\ref{fig:expfigIV}f, i. For chloride ions, the spike in the activation function sets in a stable current flow $I_\text{Cl} = b_\text{Cl} \cdot I_\text{Cl}^\text{PNP}$ in the physiological range, cf.~\cref{fig:currents}. A relatively high baseline of chloride current into the cell under physiological conditions independent of membrane potential is also noteworthy. This behavior may be due to membrane-impermeable intracellular polymer-like anionic chains that increase chloride inflow near the membrane \cite{GIANAZZA19801, glykys2014anions}. In this vicinity, which is referred to as Debye layer, electroneutrality cannot be assumed \cite{bagotsky2005fundamentals}.

The activation function of potassium bears additional remark. This function takes negative values in the physiological state, implying an inflow of potassium ions into the cell, cf.~also \cref{fig:currents}. Since the activation functions contain all ionic coupling information, this behavior can be explained by an appeal to the double-layer theory: the cell membrane is not---as assumed in our model---electrically neutral but negatively charged. Cations form a layer at close range to the negatively charged membrane, and anions accumulate in a second layer. The dominant ionic species in the first layer are potassium and calcium. Therefore, the concentrations that set the boundary conditions for both experiments and calculations can exhibit local fluctuations near the membrane. These fluctuations cannot be resolved experimentally and are neglected in calculations.

\begin{figure}[ht]
     \centering
     \includegraphics[width=\textwidth]{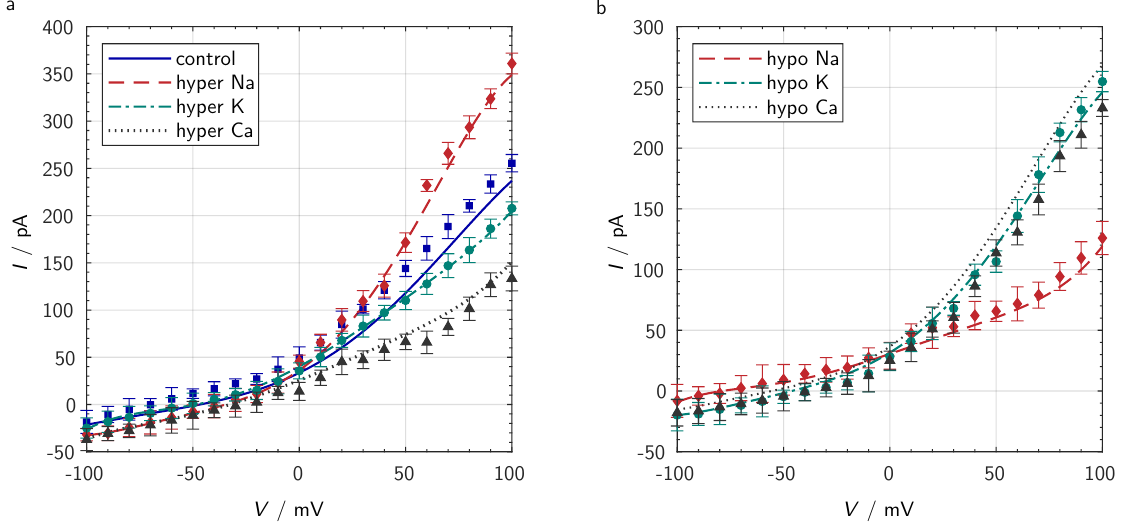}
     \caption{I/V curves of experimental data and simulation results for (a) control and hyperphysiological and (b) hypophysiological environment of the extracellular medium.}
     \label{fig:fit}
\end{figure}

The total membrane currents predicted by (\ref{eq:linearCombination}) are compared in \cref{fig:fit} against the measured I/V curves. The overall agreement with the experiment achieved by the calibrated model is remarkable, especially considering the simplicity of the model. This overall agreement notwithstanding, slight discrepancies are observed in connection with the counter-intuitive behavior of hyper- and hypophysiological environment of potassium and calcium ions for membrane potentials above 0 mV. The experimentally measured membrane current is higher for the physiological environment than for the hyper- and hypophysiological environment of potassium and calcium. In addition, both an increase and a decrease in the concentration of these two ion species lead to the same response, namely, an increase in the overall membrane current. This behavior can again be explained by an appeal to double-layer theory. 
Concentration changes of potassium or calcium do not affect the double layer, as other cations from the extracellular matrix can substitute to form a stable, positively charged layer. Consequently, a similar behavior for sodium ions is not to be expected. In our model, this behavior results in a slight underestimation of the physiological prediction and a slight overestimation of the decreased calcium prediction.

\subsection{Discussion}

\begin{figure}[ht]
    \centering
    \includegraphics[width=\textwidth]{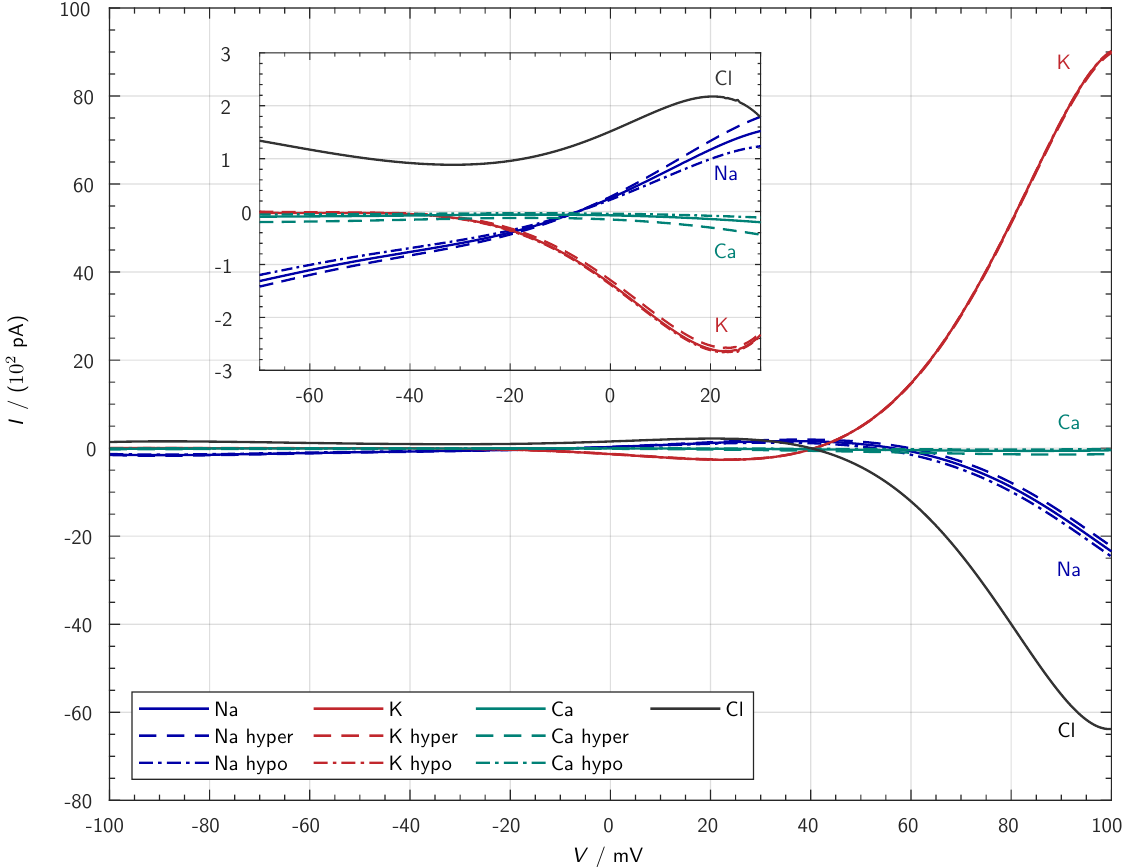}
    \caption{Ionic membrane currents of the individual ion species in experimental and physiological range.}
    \label{fig:currents}
\end{figure}

We recall that the activation functions obtained from calibration may take negative values, \cref{fig:parameters}, and, consequently, result in unexpected membrane current flow of individual ion species, as shown in \cref{fig:currents}. As already mentioned, the high baseline current of chloride ions could follow from anionic proteins that disturb the electroneutrality in short distances. However, this baseline results in a compensating inward sodium current in an equilibrium state to achieve a zero net membrane current. Impermeant anions, such as negatively charged intracellular proteins, also could be one of the main contributing causes for the high baseline of chloride ions \cite{dusterwald2018biophysical}. Consideration of constant chloride activation in computational tests evinces a resulting zero chloride current flow at equilibrium potential, which also results in membrane currents close to zero for the other ion species. This result is expected from the assumption of an anionic baseline \cite{berndt2011influence}.

Another important factor to carefully weigh is the time dependence of neuron response in experiments, cf.~\cref{fig:expfigIV}, {\sl vs.}~the stationary character of the single-channel and membrane models in the present work. In \cref{fig:expfigIV}, the measurements for hypophysiological sodium and hyperphysiological calcium environment stand out for values above $0$ mV of membrane potential. Furthermore, in the hypophysiological potassium environment, a shorter response time is observed for each of the measurements. Overall, the close agreement between steady-state PNP and experiments suggests that the time scale of any transient effects is negligibly small and appears to bear out the assumption of stationary channel flow. 
\section{Concluding remarks}
\label{sec:5conclusion}

Realistic models that accurately represent anatomical detail and the mechanical response of the tissues in the human skull are available from Magnetic Resonance Imaging (MRI) \cite{weickenmeier2017a, lu2019a, ganpule2017a, miller1998a}, Magnetic Resonance Elastography (MRE) \cite{ophir1991a, nightingale2003a, Lan:2020, Weickenmeier:2018}, and other imaging techniques, which enables finite-element analyses of wave propagation in the brain under a variety of conditions from concussion to ultrasound neuromodulation, cf., e.~g., \cite{sayed2008a, Budday:2020, salahshoor2020a}. By contrast, there is a paucity of neuronal ion-channel and cell-membrane models that are quantitatively predictive and can be integrated into the full-scale finite-element analyses to predict the extent of neuronal activation as a function of local conditions. Indeed, not until the recent breakthrough work of Yoo {\sl et al.} \cite{yoo2022} the precise mechanisms by which local strain and ultrasound activate neural activity have not been conclusively known.

The present work is intended as a first step in filling in this modeling gap. We have shown that the combination of simplified PNP calculations and ion-specific activation functions calibrated from experimental data accurately predicts ionic currents and I/V curves as a function over the entire physiological range of membrane potentials. A number of aspects of the membrane response, such as the dynamic response of voltage-gated ion channels, the continuous effect of the sodium-potassium ATPase pump, calcium-dependent signaling pathways, and electrodiffusion phenomena in the Debye layer, especially the double layer and negatively charged intracellular proteins \cite{savtchenko2017electrodiffusion}, complex ion channel geometries and intracellular biochemical interactions such as calcium-based intracellular signaling pathways \cite{kefauver2020discoveries, grienberger2012imaging}, are not accounted for explicitly by the model but only implicitly, if at all, by calibration to the experimental data. It is conceivable that an explicit accounting of these and other effects could improve the predictiveness of the model without incurring excessive computational complexity. These and other enhancements of the model suggest themselves as worthwhile directions for further research.

\section*{Data availability statement}

Data and code (programmed in MATLAB \cite{MATLAB}) for the single-channel model and the full-axon transport model are available at DaRUS \cite{darus}.

\section*{Acknowledgements}

This work is funded by the German Research Foundation (Deutsche Forschungsgemeinschaft; DFG) within the Priority Program 2311, grant number 465194077, and the Max Planck Society. We furthermore gratefully acknowledge the support of the DFG under Germany’s Excellence Strategy -- EXC 2075 -- 390740016. E.Y.\ has received funding from the European Union’s Horizon 2020 research and innovation program under the Marie Sk\l{}odowska-Curie grant agreement no 101059593.

\end{document}